\documentclass[a4paper,10pt,twoside]{cpc-hepnp}

\usepackage{multicol}
\usepackage{graphicx}
\usepackage{booktabs}
\usepackage{amssymb,bm,mathrsfs,bbm,amscd}
\usepackage[tbtags]{amsmath}
\usepackage{lastpage}
\usepackage[encapsulated]{CJK}

\def \bqn {\begin{equation}}
\def \eqn {\end{equation}}
\def \bqna {\begin{eqnarray}}
\def \eqna {\end{eqnarray}}

\def \ee {e^+e^-}

\begin{document}

\fancyhead[c]{\small Chinese Physics C~~~Vol. xx, No. x (2015) xxxxxx}
\fancyfoot[C]{\small 010201-\thepage}

\footnotetext[0]{Received xx May 2015}

\title{Interference effects on Higgs mass measurement in $e^+e^-\to H(\gamma\gamma) Z$ at CEPC
\thanks{supported by the National Natural Science
Foundation of China (Grants No. 11375021 and No. 11447018), the New Century Excellent Talents in University (NCET) under grant
NCET-13-0030,  the Major State Basic Research Development Program of China (No. 2015CB856701), the Fundamental Research Funds for the Central Universities, and the Education Ministry of
LiaoNing Province.
}}

\author{
Xu Guang-Zhi
$^{1}$
\quad Li Gang$^{2}$
\quad Li Yi-Jie$^{3}$
\quad Liu Kui-Yong$^{3}$
\quad Zhang Yu-Jie$^{1,4;1}$\email{nophy0@gmail.com}
}
\maketitle

\address{%
$^1$ School of Physics,  Beihang University, Beijing 100191, China\\
$^2$ Institute of High Energy Physics, Chinese Academy of Sciences, Beijing 100049, China\\
$^3$ Department of Physics, Liaoning University, Shenyang 110036, China\\
$^4$ CAS Center for Excellence in Particle Physics, Beijing 100049, China
}

\begin{abstract}
A high luminosity Circular Electron Positron Collider (CEPC)
as a Higgs Factory will be helpful to the precision  measurement of the Higgs mass.
The signal-background interference effect is carefully studied for the Higgs diphoton decay mode
in the associated Z boson production at the future $e^+e^-$ colliders at energy  $246 {\rm GeV}$.
The mass shifts go up from about $20 {\rm MeV}$ to $50  {\rm MeV}$ for the experimental mass resolution ranging from $0.8 {\rm GeV}$ to $2 {\rm GeV}$.
\end{abstract}


\begin{pacs}
1---3 PACS(Physics and Astronomy Classification Scheme, http://www.aip.org/pacs/pacs.html/)
\end{pacs}

\footnotetext[0]{\hspace*{-3mm}\raisebox{0.3ex}{$\scriptstyle\copyright$}2015
Chinese Physical Society and the Institute of High Energy Physics
of the Chinese Academy of Sciences and the Institute
of Modern Physics of the Chinese Academy of Sciences and IOP Publishing Ltd}%

\begin{multicols}{2}

\section{Introduction}

The ATLAS and CMS collaborations at the Large Hadron Collider (LHC) announced the amazing discovery of a new particle
with a mass of around $125 {\rm GeV}$ in July 2012\cite{Chatrchyan:2012ufa,Aad:2012tfa}
and it's properties are well compatible with the Standard Model (SM) Higgs boson\cite{Dawson:2013bba} but leave the room for new physics.
One of next  tasks is the precision measurement, such as the mass, spin, couplings, and decay patterns, to determine the nature for Higgs boson.
The future $e^+e^-$ colliders such as the International Linear Collider (ILC),
a linear particle accelerator, Triple-Large Electron-Positron Collider (TLEP),
and the Circular Electron Positron Collider (CEPC),
proposed by Chinese high energy physics community in 2012, will play an important role on this task.

For future Higgs factories, the $e^+e^-\to Z^\ast\to Z+H(\gamma\gamma)$ process will be an excellent channel for precision measurement.
The diphoton decay channel is a rare decay mode but provides a clear signature for the Higgs boson.
At  the LHC, the $\gamma\gamma$ mode for Higgs production involves a huge background including the dominant reducible jet and the irreducible contributions from the continuum.
The background is significantly suppressed at lepton colliders to measure the Higgs properties\cite{Mo:2015mza}.
The Higgs-bremsstrahlung process, $e^+e^-\to Z H$, is the most important process for Higgs production
when the center-of-mass energy is less than $500 {\rm GeV}$.
With the leading-order calculations, the production cross section of a $125 {\rm GeV}$ Higgs reaches the maximum value when the center-of-mass energy is around $246 {\rm GeV}$.

The diphoton decay rate has been calculated up-to the complete three-loop level\cite{Maierhofer:2012vv}
and a four-loop estimation is also considered\cite{Sturm:2014nva}.
The contributions at three-loop and four-loop levels can be neglected in comparison of the one-loop decay rate.
For the two-loop level, the QCD and electroweak corrections are nearly completely cancelled in the numerical calculations
for Higgs with mass of $125 {\rm GeV}$. 
The electroweak radiative correction for $e^+e^- \to Z H$ was calculated\cite{Denner:1991ue,Denner:1992bc,Englert:2013tya}
and the contribution is only below $5\%$ of the tree-level cross section for Higgs with mass of $125 {\rm GeV}$\cite{Englert:2013tya}.
For the background from the continuum, the next-to-leading order electroweak corrections have been considered
for the $e^+e^-\to Z \gamma\gamma$ process in the SM by Y. Zhang et. al in the recent work\cite{Yu2014}.
A correction of $2.32\%$ is observed as the center-of-mass energy increased to $250 {\rm GeV}$.  Though the investigation of several typical distributions for the final photons, they also found a dramatical separation in background and in the signal process,
which indicated that the background can be significantly suppressed to study the Higgs signal by taking appropriate kinematic cut. 

The interference effects for the Higgs mass in the diphoton decay mode at the hadron colliders have been discussed based on the theoretical aspects.
The signal-continuum interference for diphoton final states at LHC was first studied by L. J. Dixon and M. S. Siu in 2003\cite{Dixon:2003yb}.
According to S. P. Martin, the mass shift from the interference effect is $150 {\rm MeV}$ or more\cite{Martin:2012xc},
but the effect becomes smaller on final states containing one extra jet\cite{Martin:2013ula}.
The interference was also evaluated to the next-to-leading order level in recent works\cite{Dixon:2013haa,deFlorian:2013psa}.
The interference effect of other final states at hadron colliders was also considered in Refs.
\cite{Campbell:2013una,Campbell:2013wga,Campbell:2011cu,Campbell:2014gua,Niezurawski:2002jx,Morris:1993bx,Dicus:1994bm}.

This work will focuse on the interference effect of the Higgs mass through diphoton decay mode in Higgs-bremsstrahlung process at the future $e^+e^-$ colliders CEPC (Several discussions at CEPC refer to Refs.\cite{Xiu:2015tha, Ruan:2014xxa, Fan:2014vta, Zhang:2014eqa}).
Recently, this interference effect with fixed polarisation at the initial state has been considered in Ref.\cite{Liebler:2015eja},
and a mass shift in range of $\mathcal{O}(100{\rm MeV})$ is found.
In this work, the interference effect but with unpolarized initial state will be revisited.

\section{Calculations and analysis}

Fig.\ref{fig:feyndia} shows the typical Feynman diagrams for the calculation of interference contributions.
Higgs boson has a very narrow width.
\begin{center}
\includegraphics[width=0.3\textwidth]{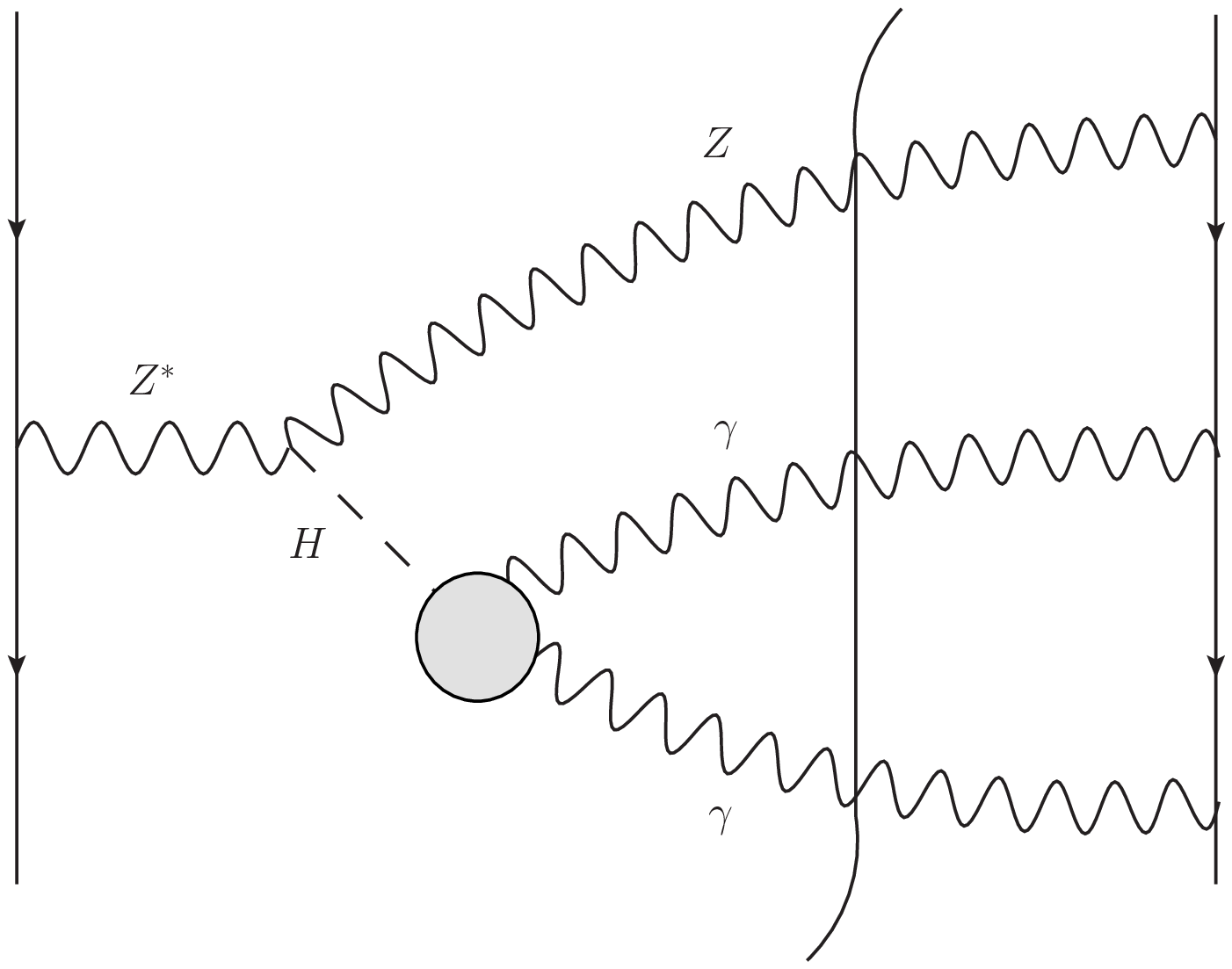}
\figcaption{\label{fig:feyndia} Typical Feynman diagrams for the interference with the continuum.  }
\end{center}
Following the method used in Refs.\cite{Dixon:2013haa,deFlorian:2013psa,Martin:2013ula,Martin:2012xc,Dixon:2003yb},
with the narrow-width approximation, the pure signal and interference cross sections for the production can be expressed as:
\bqna
&&\frac{d\sigma^{sig}}{dM_{\gamma\gamma}}=\frac{\big|\mathcal{A}_{\ee\to ZH}\mathcal{A}_{H\to\gamma\gamma}\big|^2}{(m^2_{\gamma\gamma} - m^2_H)^2+m^2_H\Gamma^2_H},\nonumber\\
&&\frac{d\sigma^{int}}{dM_{\gamma\gamma}}=\frac{-2(m^2_{\gamma\gamma} - m^2_H)R-2m_H\Gamma_H I}{(m^2_{\gamma\gamma} - m^2_H)^2+m^2_H\Gamma^2_H},
\eqna
where $R$ and $I$ represent the real and imaginary parts of the interference amplitude
$(\mathcal{A}_{\ee\to ZH}\mathcal{A}_{H\to\gamma\gamma}\mathcal{A}^\ast_{cont})$, respectively,
and $\mathcal{A}^\ast_{cont}$ is the continuum amplitude.
The real part is odd in the vicinity of Higgs mass because of the factor $m^2_{\gamma\gamma} - m^2_H$
and the total contribution to the decay width is negligible.
However, as stated in Ref.\cite{Martin:2012xc}, a sharp peak and a dip exist near the $M_H$ in the diphoton distribution,
and the effect slightly moves the peak position.
As mentioned in the introduction, the next-to-leading order electroweak corrections
to the continuum part only contribute less than $5\%$ to the tree-level cross section.
Therefore, only the tree-level contributions for the continuum part is considered in our calculation.
For the amplitude of Higgs boson coupled with two photons, we also apply
the result at one-loop level\cite{Dixon:2013haa,Martin:2013ula,Martin:2012xc,Dixon:2003yb}.
For the input parameters, the resonance mass and width of $M_H=125.6{\rm GeV}$,
$\Gamma_H=4.2{\rm MeV}$, the fine structure constant  $\alpha=1/137$,
and the running fermion masses $m_t=168.2{\rm GeV}$, $m_b=2.78{\rm GeV}$, $m_c=0.72{\rm GeV}$, $m_{\tau}=1.744{\rm GeV}$ are adopted, respectively.
The signal cross section would reach the maximum at round $245-246 {\rm GeV}$.
Here we take the center-of-mass energy of $246 {\rm GeV}$ in the following calculations.

Fig.\ref{fig:sig_con} illustrates the pure signal for diphoton production from Higgs decay and the continuum cross sections.
\end{multicols}
\ruleup
\begin{center}
 \includegraphics[width=0.3\textwidth]{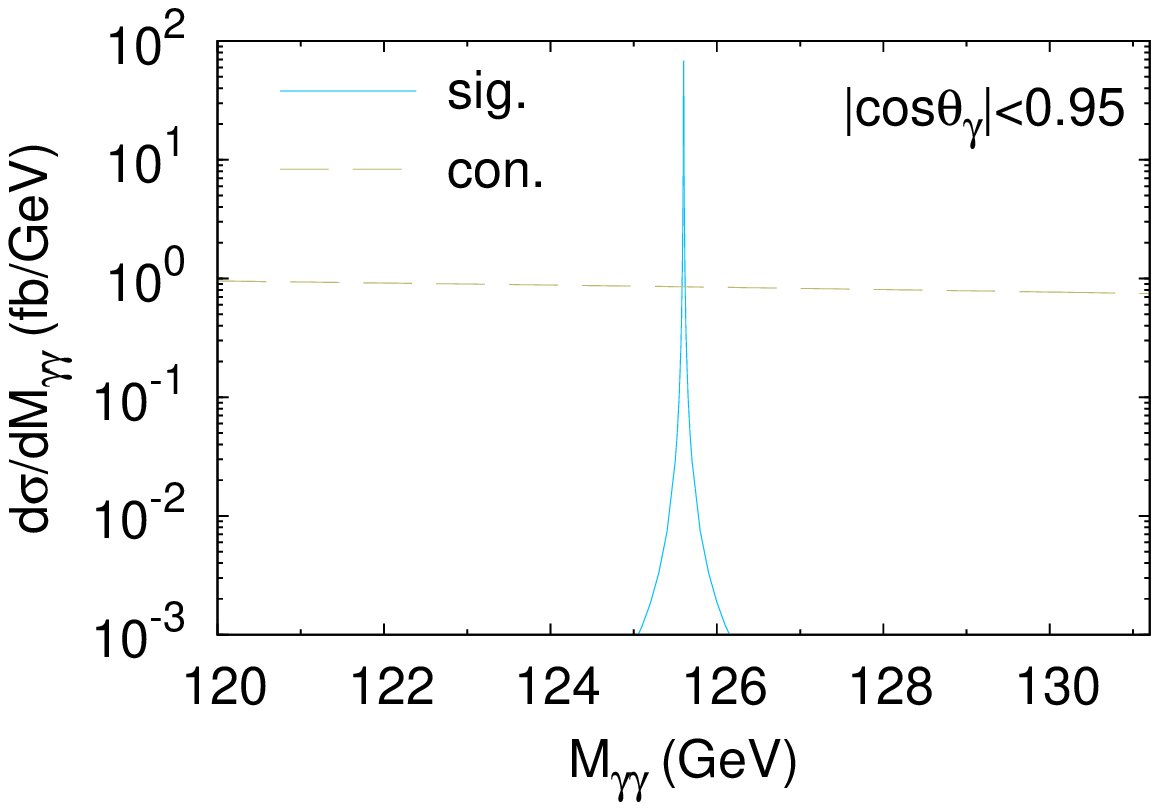}
  \includegraphics[width=0.3\textwidth]{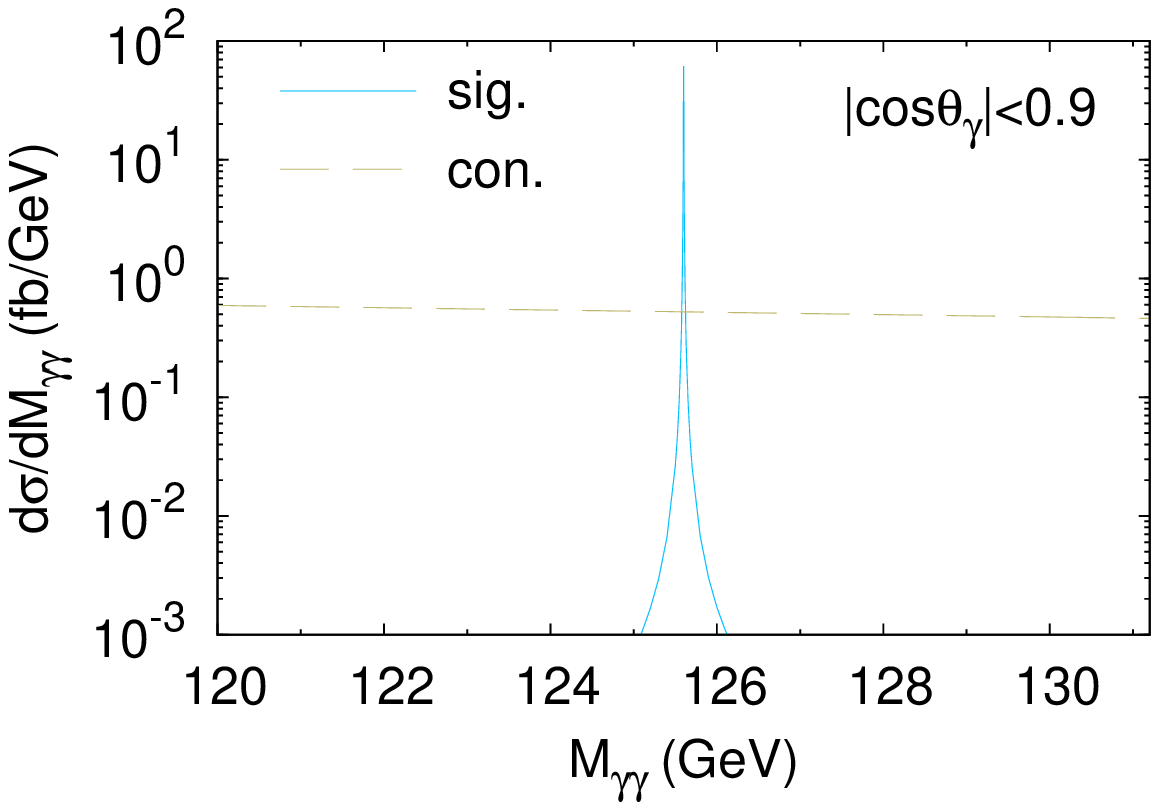}
   \includegraphics[width=0.3\textwidth]{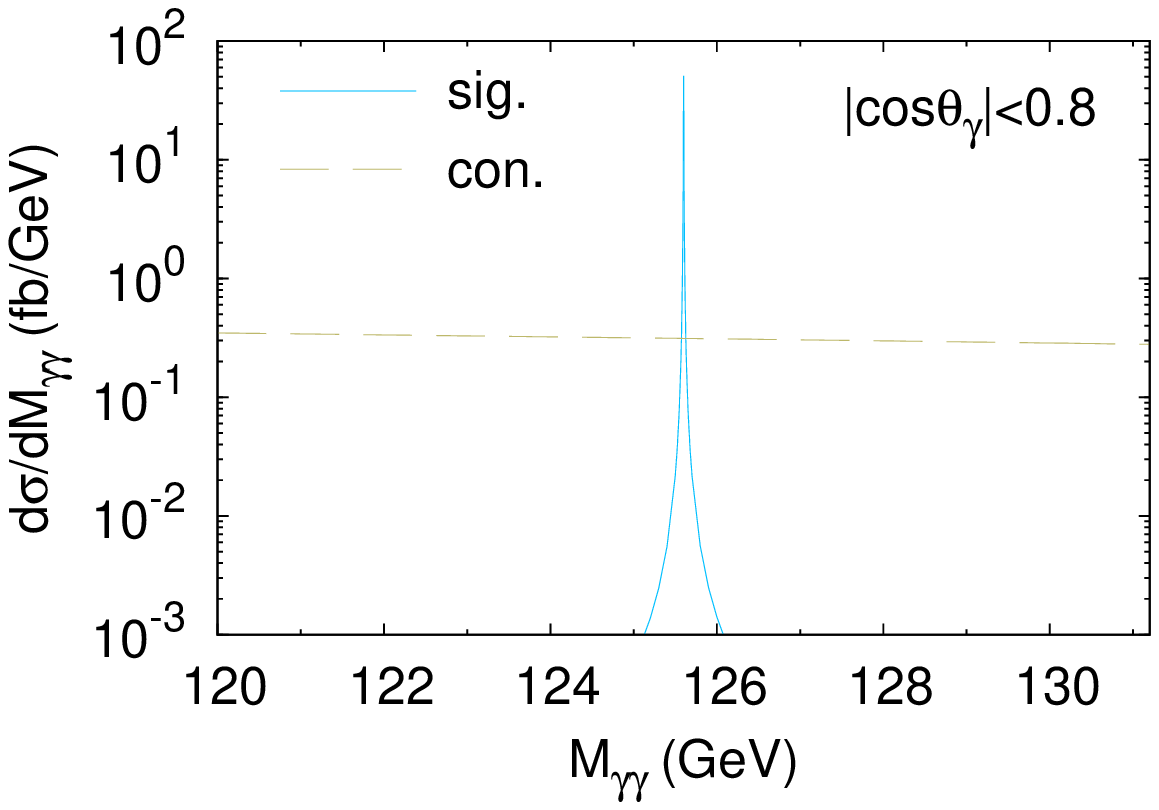}
\figcaption{\label{fig:sig_con} (color online) Comparison for the background and signal process with different cut conditions for the final photons. }
\end{center}
\ruledown
\begin{multicols}{2}
Three type of cuts on the scattering angle are implemented, which are
$\vert {\rm cos}\theta^{cut}_{\gamma}\vert =0.8$, $\vert {\rm cos}\theta^{cut}_{\gamma}\vert =0.9$
and $\vert {\rm cos}\theta^{cut}_{\gamma}\vert =0.95$, respectively.
Notably, the experimental cut for the scattering angle may be larger than these values.
However, the cross section from ISR, i.e. continuum contribution, is sensitive to the choice of this kind of cuts
because its behavior depends on $1/(1-{\rm cos}\theta^{cut}_{\gamma})$.
That indicates the background contributions sharply increase compared with the signals when larger angle cuts are chosen,
making the interference effect to the mass measurement insensitive to a larger cut.
Another cut on the final photon energy is taken to $20 {\rm GeV}$.
For the above cut selection, the signal process has sharp peak at the range of $50-70 {\rm fb}$,
and the continuum cross sections only reaches $0.5-1 {\rm fb}$ with the diphoton invariant mass in the range of $120-130 {\rm GeV}$.

The real-part cross sections of the interference as a function of the diphoton invariant mass are shown on the left panel of Fig.\ref{fig:res_re_int},
whereas the signal with and without the interference effect are shown on the right panel of Fig.\ref{fig:res_re_int}.
\end{multicols}
\ruleup
\begin{center}
\includegraphics[width=0.4\textwidth]{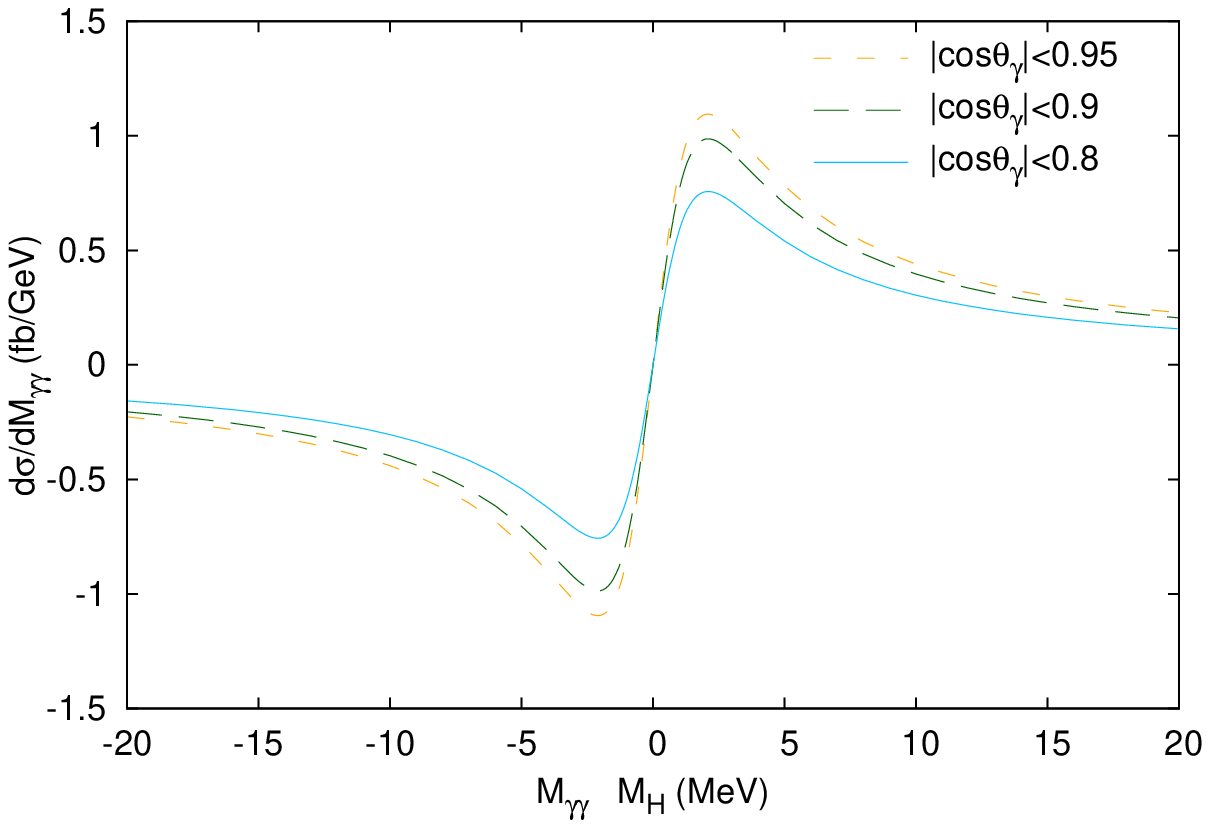}
\includegraphics[width=0.4\textwidth]{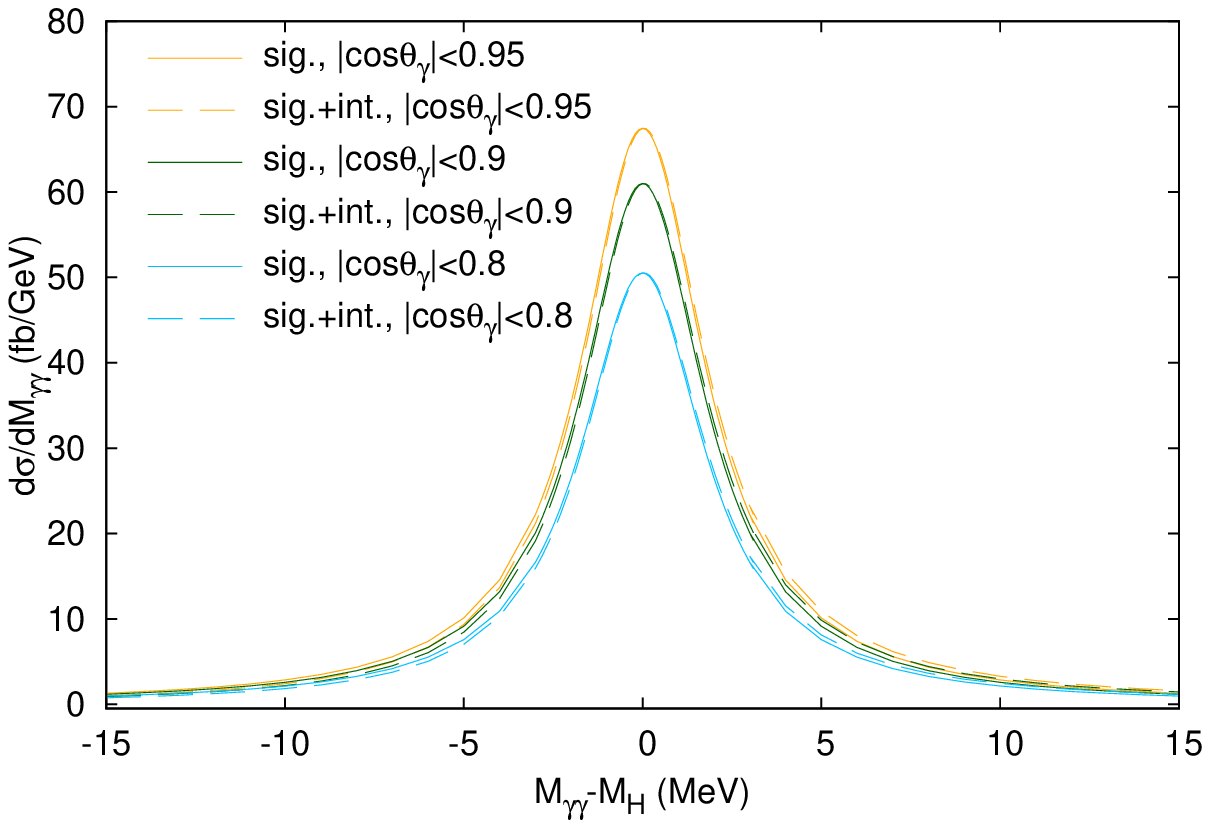}
\figcaption{\label{fig:res_re_int} (color online) The left panel shows the diphoton invariant mass distribution from the real interference.
	The right shows the signals with and without the interference from the background.
	The cut of scattering angle for the photons is chosen as $\vert \cos\theta_{\gamma}\vert <0.8$, $\vert \cos\theta_{\gamma}\vert<0.9$
	and $\vert \cos\theta_{\gamma}\vert <0.95$.
   The cut of the final photon energy is $E_{\gamma}>20 {\rm GeV}$.}
\end{center}
\ruledown
\begin{multicols}{2}
The imaginary part effects are observed to be even but negligible negative values with their maximum values are less than $2\%$ of that of real part cross sections, and we neglect then in the analysis.

To study the effect of the interference to the Higgs mass measurement, in Ref. \cite{Martin:2012xc},
convolution integrals with a Gaussian function were created to the cross section
to simulate the smearing effect of the Higgs mass due to finite experimental resolution.
In Fig.\ref{fig:sig_int}, the results are plotted with the Gaussian width as $\sigma_{MR}=0.8$, $1.0$, $1.5$, and $2.0 {\rm GeV}$.
Compared with the signals without smearing effects shown in Fig.\ref{fig:res_re_int},
the peak slightly moves toward the larger mass direction when the interference effects are taken into account.
The behavior of the right-side shift is consistent with that shown in \cite{Liebler:2015eja}.
Similar effects are also observed in the previous studies of hadron colliders
\cite{Dixon:2013haa,deFlorian:2013psa,Martin:2013ula,Martin:2012xc,Dixon:2003yb}
(The left-side or right-side shift effect might occurs for different sub-process).
\end{multicols}
\ruleup
\begin{center}
\includegraphics[width=0.4\textwidth]{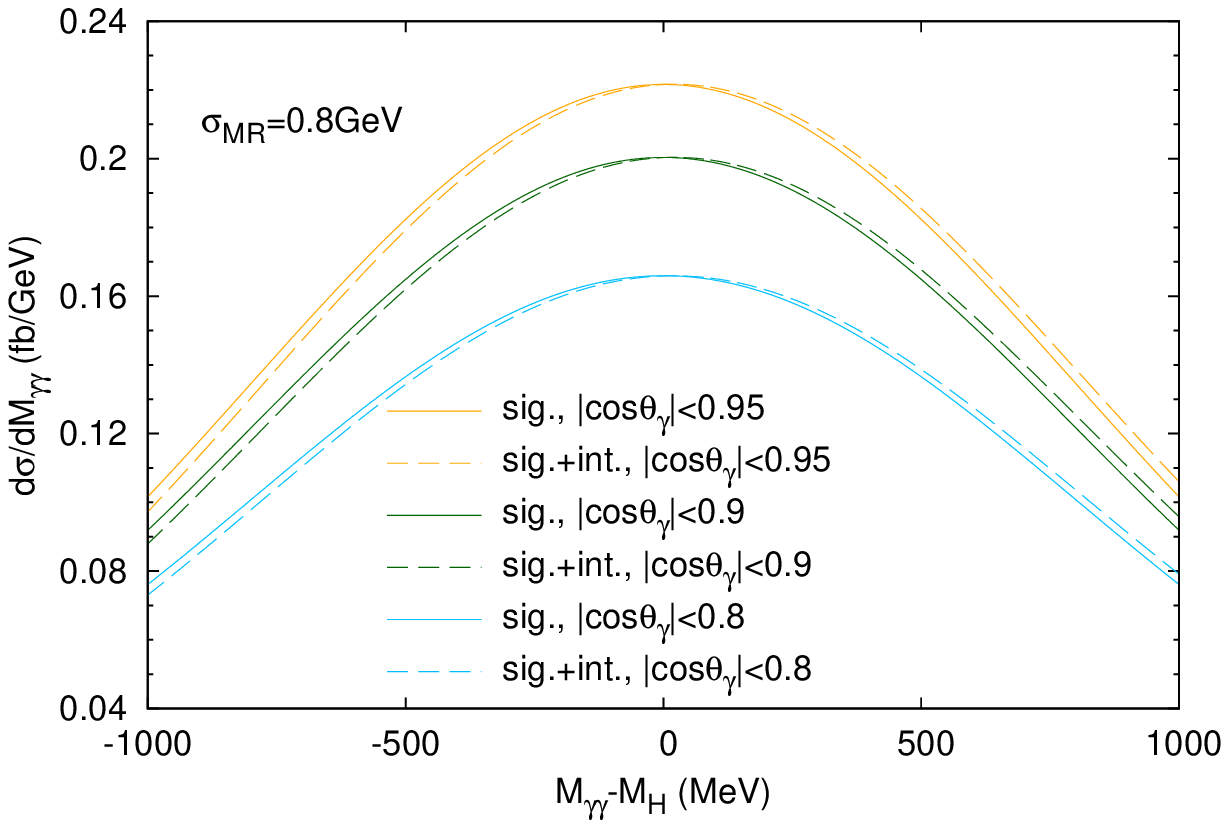}
\includegraphics[width=0.4\textwidth]{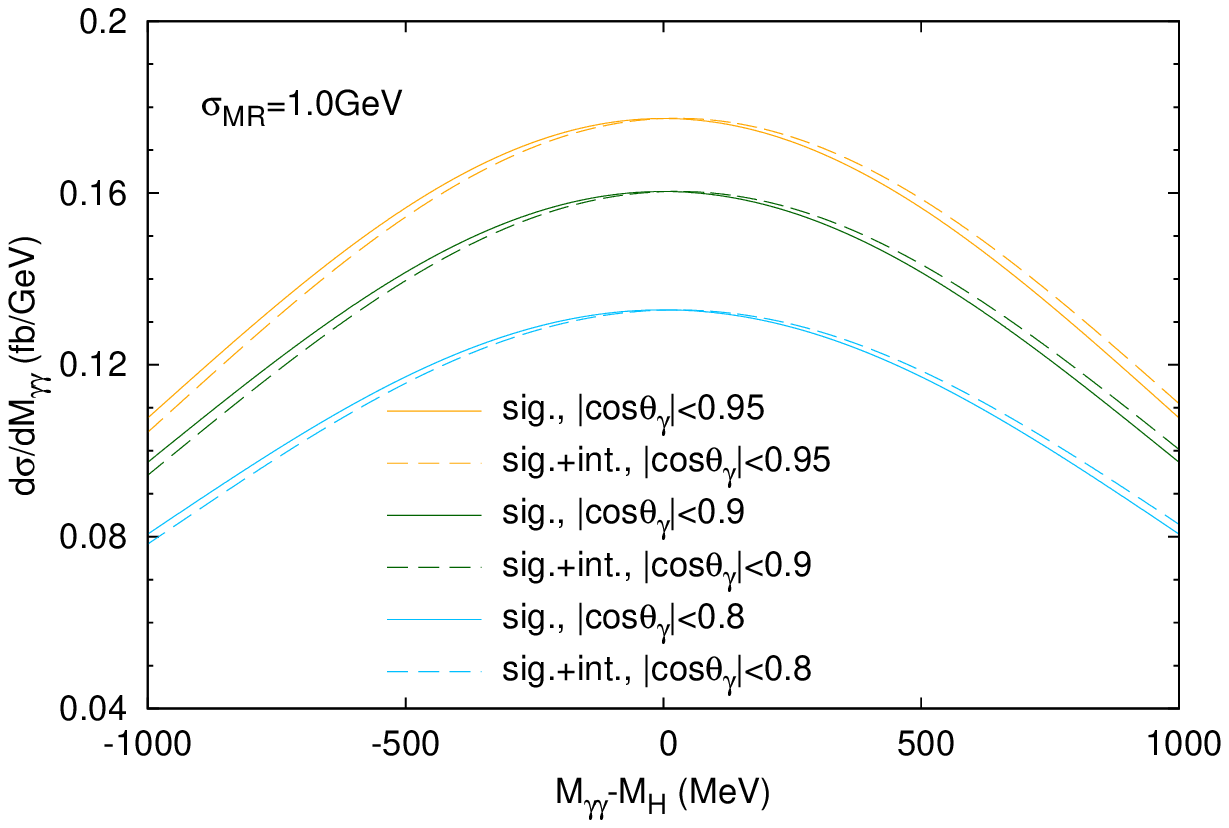}\\
\includegraphics[width=0.4\textwidth]{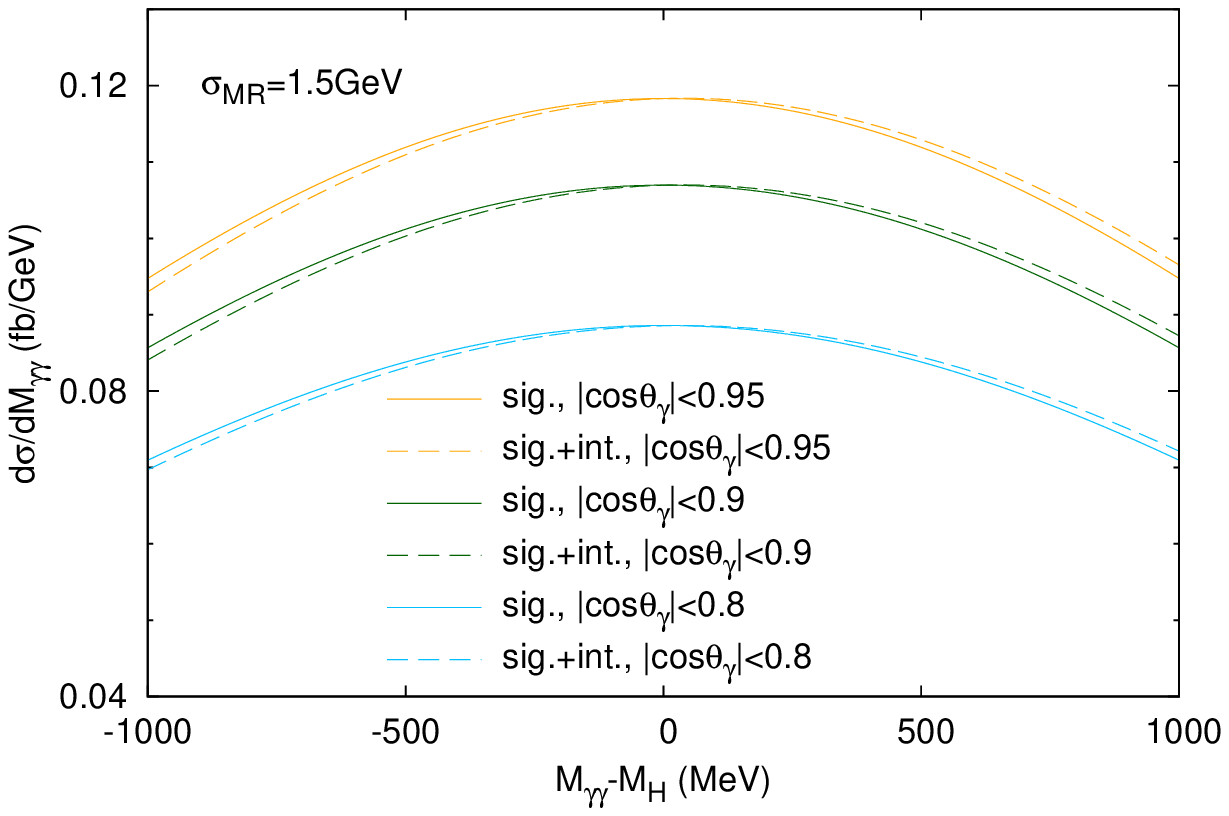}
\includegraphics[width=0.4\textwidth]{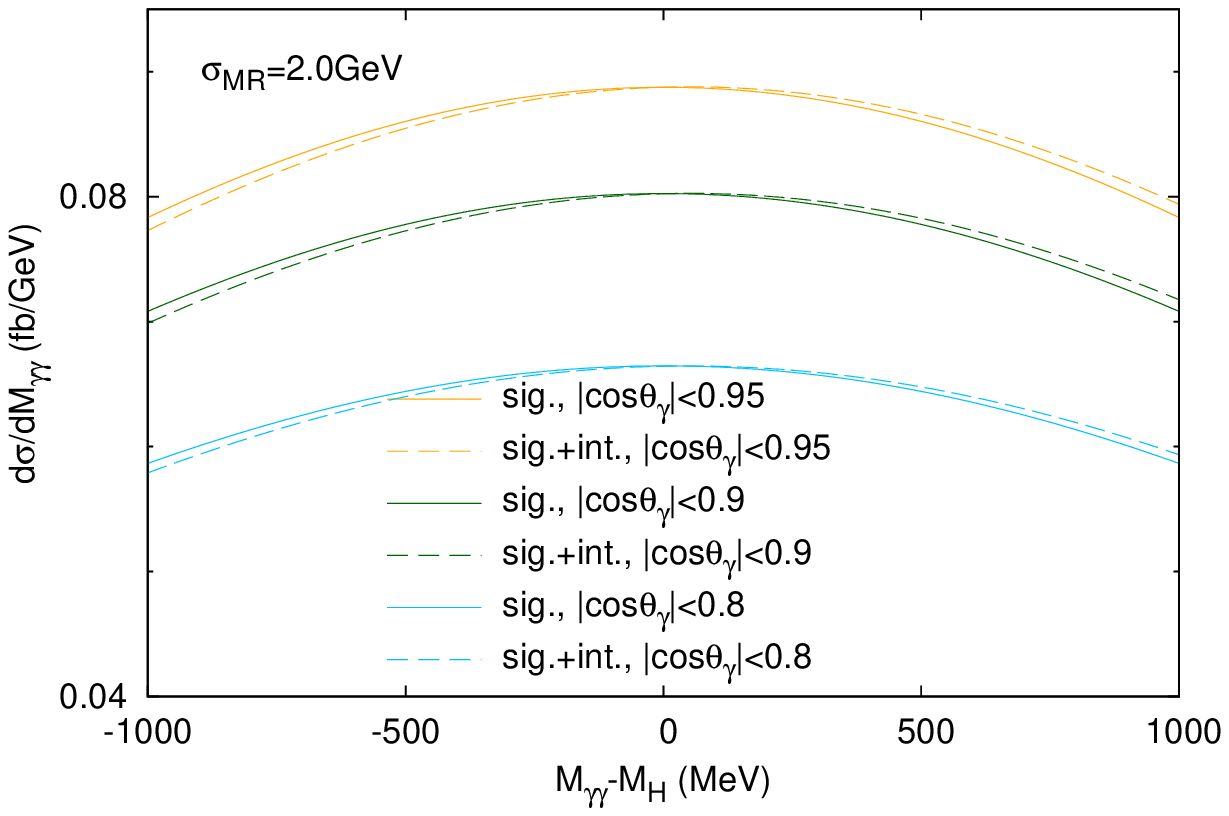}
\figcaption{\label{fig:sig_int}
	(color online) Diphoton invariant mass distributions of Higgs signal with different mass resolutions and kinematic cuts.
   The input parameters, mass resolutions ($\sigma_{MR}$) and cut of scattering angle for the photons
	are noted in the plots. The cut of the final photon energy is $E_{\gamma}>20 {\rm GeV}$.}
\end{center}
\ruledown
\begin{multicols}{2}

The strategy stated by S. P. Martin in Ref.\cite{Martin:2013ula}
is applied to estimate the mass shift
and a least-square fit to the line shape of mass shifts as a function of the Gaussian width ($\sigma_{MR}$) is performed.
The results are shown as Fig.\ref{fig:res_massshift}.
\begin{center}
\includegraphics[width=0.48\textwidth]{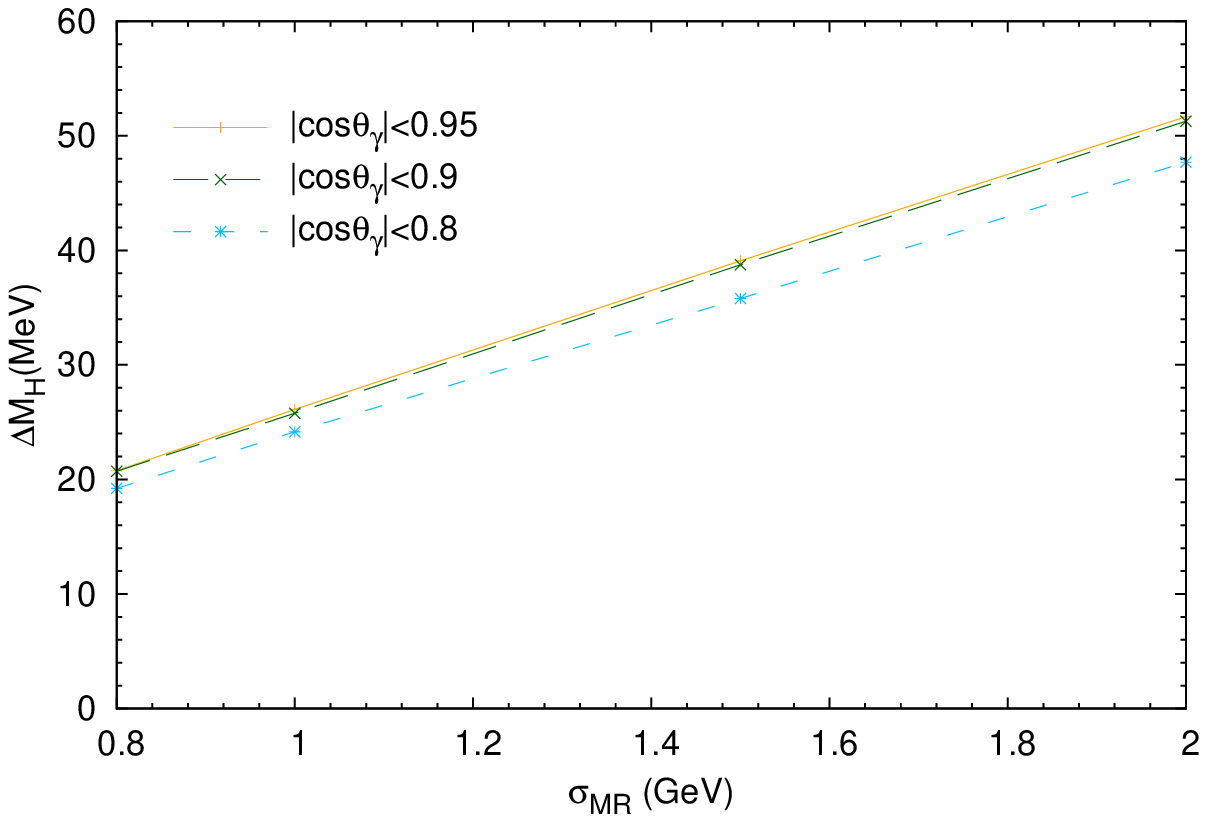}
\figcaption{\label{fig:res_massshift}
	(color online) The Higgs mass shifts due to the signal-background interference as a function of the Gaussian mass resolution width.
	Here, the least-square fit to the line shape of mass shifts is performed according to the strategy proposed by S. P. Martin\cite{Martin:2013ula} .}
\end{center}

Similar to the case of Higgs production in the associated one jet,
the mass shifts linearly increased with increasing mass resolution width $\sigma_{MR}$\cite{Martin:2012xc}.
The mass shifts increased from about $20 {\rm MeV}$ to $50 {\rm MeV}$,
corresponding to the range of the mass solution width from $0.8 {\rm GeV}$ to $2.0 {\rm GeV}$.
Form the figure, the two lines corresponding to the final photon scattering angle cut ${\rm |cos\theta^{cut}_\gamma|=0.9}$ and ${\rm |cos\theta^{cut}_\gamma|=0.95}$, respectively, are very close to each other referring to the gap with the line ``${\rm |cos\theta^{cut}_\gamma|=0.8}$".
This result implies that the shifts from the interference effect are not sensitive to a larger angle cut, as mentioned in the above analysis.

\section{ Summary}

Therefore, in this work, followed by the previous works regarding hadron colliders
\cite{Dixon:2013haa,deFlorian:2013psa,Martin:2013ula,Martin:2012xc,Dixon:2003yb},
the signal-background interference effect of the Higgs mass through diphoton decay mode
in the associated Z boson production at the future $e^+e^-$ colliders at energy $240\sim250 {\rm GeV}$ was considered.
Different cut conditions for the final photon scattering angle and different smearing width to simulate the experiments were also considered.
Similar right-side shifts to the Higgs mass in the spectrum were also observed,
consistent with the statement in Ref.\cite{Liebler:2015eja}.
Considering the smearing Gaussian width $\sigma_{MR}$ (which simulated to the experimental mass resolution)
ranging from $0.8 {\rm GeV}$ to $2 {\rm GeV}$,
the corresponding mass shifts increased from about $20 {\rm MeV}$ to $50  {\rm MeV}$.
These results will be beneficial in studying the precision measurement of the Higgs mass.


\end{multicols}

\vspace{15mm}

\vspace{-1mm}
\centerline{\rule{80mm}{0.1pt}}
\vspace{2mm}

\begin{multicols}{2}


\end{multicols}

\clearpage

\end{document}